# Deterministic integration of single nitrogen-vacancy centers into nanopatch antennas

S.I. Bogdanov[1,2][*], O.A. Makarova[1,2][*], A.S. Lagutchev[1,2], D. Shah[1,2], C.C. Chiang[1,2], S. Saha[1,2], A.S. Baburin[3,4], I.A. Ryzhikov[3,5], I.A. Rodionov[3,4], A.V. Kildishev[1,2], A. Boltasseva[1,2] and V.M. Shalaev[1,2]

[1] School of Electrical & Computer Engineering and Birck Nanotechnology Center, Purdue University, West Lafayette, Indiana 47907, United States
[2] Purdue Quantum Science and Engineering Institute, Purdue University, West Lafayette, Indiana 47907, United States
[3] FMNS REC, Bauman Moscow State Technical University, Moscow 105005, Russia
[4] Dukhov Research Institute of Automatics, Moscow 127055, Russia
[5] Institute for Theoretical and Applied Electromagnetics RAS, Moscow 125412, Russia

**Abstract:**

Quantum emitters coupled to plasmonic nanoantennas produce single photons at unprecedentedly high rates in ambient conditions. This enhancement of quantum emitters' radiation rate is based on the existence of optical modes with highly sub-diffraction volumes supported by plasmonic gap-nanoantennas. Nanoantennas with gap sizes on the order of few nanometers have been typically produced using various self-assembly or random assembly techniques. Yet, the difficulty of controllably fabricate nanoantennas with the smallest mode sizes coupled to pre-characterized single emitters until now has remained a serious issue plaguing the development of quantum plasmonic devices. We demonstrate the transfer of nanodiamonds with single nitrogen-vacancy (NV) centers to an epitaxial silver substrate and their subsequent deterministic coupling to plasmonic gap nanoantennas. Through fine control of the assembled nanoantenna geometry, a dramatic shortening of the NV fluorescence lifetime was achieved. We furthermore show that by preselecting NV centers exhibiting a photostable spin contrast, a coherent spin dynamics can be measured in the coupled configuration. The demonstrated approach opens unique applications of plasmon-enhanced quantum emitters for integrated quantum information and sensing devices.

## Introduction

Solid-state optical quantum emitters such as quantum dots, color centers, and rare earth ion impurities are fueling the rise of quantum information and sensing technologies[1]. Quantum emitters can produce indistinguishable single photons on demand[2–4], realize interactions between single photons[5] as well as store and process quantum information in the highly protected spin states[6–8]. They can also be used to sense electromagnetic fields[9,10], temperature[11], and strain[12] at the nanoscale. In order to take full advantage of their varied and unique functionalities, one must carefully engineer the interaction of these emitters with electromagnetic fields. Advanced engineering of the optical radiative spontaneous decay is of particularly importance for emerging applications. A variety of structures based on dielectric, plasmonic, and low-index materials has been proposed and demonstrated to speed-up the spontaneous emission process and increase the emitter brightness[13–15].

In this regard, plasmonic nanoantennas[14] are a unique class of nanostructures because they offer the highest single-photon rate enhancement factors[16]. This enhancement is proportional to the ratio of the mode quality factor $Q$ to its volume $V$. In dielectric resonators, as $Q$ increases, the radiative quantum emission rate becomes eventually limited by the emitter's intrinsic optical linewidth or by the strong interaction limit. On the other hand, $V$ in such resonators is diffraction-limited. In plasmonic nanoantennas, V can be several orders of magnitude below the diffraction limit. At the same time, plasmonic modes are broadband with quality factors ranging from $Q = 10^1$ to $10^3$, which allow for the enhancement of relatively broadband emitters. Such broadband operation may for example alleviate the issues with frequency tuning, commonly encountered in the experiments with dissimilar emitters[17,18] or high-$Q$ resonators[19]. On the other hand, plasmonic enhancement factors are highly sensitive to the nanoscale details of the device geometry. In particular, at visible frequencies, only gap sizes below or about 20 nm lead to a significant fluorescence lifetime shortening (100 times and more)[20]. As a consequence, single emitter coupling to highly sub-diffraction volume modes in plasmonic nanoantennas (thus forming single-photon nanoantennas) is challenging to achieve using standard nanofabrication methods. In addition, processing steps such as lithography, etching and milling may introduce unwanted contaminants which can create a strong emission background. Therefore, new fabrication methods need to be established for the realization of nanoscale quantum plasmonic devices. A variety of self-assembly methods[21–23] or random assembly methods[24] have been employed for plasmonic gap antennas, achieving gaps below 20 nm. However, a fully deterministic fabrication, i.e. whereby a characterized emitter is coupled to an antenna with a highly controlled geometry remains one of the field's key challenges[14,25].

---

[*] These authors contributed equally to this work

Scanning-probe microscopy has been considered as an increasingly viable tool for the assembly of quantum devices from individual constituents[26,27]. One can pick, place, and move nanoparticles on substrates with high precision using a tip of an atomic force microscope (AFM). However, as it comes to single-photon nanoantennas, this method has until now been limited to obtaining gaps of about 30 nm and wider[28–30]. Such gaps are still too large to reap fully the benefits of plasmonic enhancement. In this work, we demonstrate a plasmonic nano-patch antenna (NPA), with gap sizes below 20 nm, deterministically assembled around a pre-characterized NV center in a nanodiamond (ND). We first numerically study the fluorescence lifetime shortening (FLS) as a function of the relative nanoparticle positions and experimentally realize the configuration with a measured FLS of over 3000 times, achieving the shortest fluorescence lifetime in NV centers to date. Furthermore, by choosing an emitter with a robust spin-dependent fluorescence signal, we demonstrate that the NV spin coherence time is of the same order in the coupled and uncoupled configurations. We discuss how the NV spin coherence in NPAs can be utilized for novel quantum information and sensing devices.

### **Nanoantenna assembly:**

NPAs are a highly attractive type of plasmonic antennas consisting of individual metal nanocubes separated from a flat metal film by a thin dielectric spacer layer. Similar to other nanoparticle-on-metal systems[31], NPAs offer modes with volumes in the range of $10^{-4} \lambda_0^3$ to $10^{-6} \lambda_0^3$ that can be finely controlled by the thickness of the spacer layer. When coupled to quantum emitters, NPAs have already been shown to provide significant FLS values[24] and ultrabright antibunched emission[32] at room temperature.

The assembly process is conducted in two stages. In the first stage, nanodiamonds are deposited on an auxiliary coverslip substrate, suitable nanodiamonds are identified and transferred onto an epitaxial silver film (Fig. 1a). In the second stage, Ag nanocubes are randomly deposited and deterministically coupled to the NV centers using an AFM tip (Fig. 1b). The characterization of the assembled NPA is conducted through an air objective with NA = 0.9. The spin state is controlled by a microwave signal using a proximal Cu wire (Fig. 1c). This two-stage process forgoes the optical characterization of emitters in nanoparticles directly dispersed on a flat metal film. Such emitters couple poorly into the far field and cannot be optically characterized in an efficient manner.

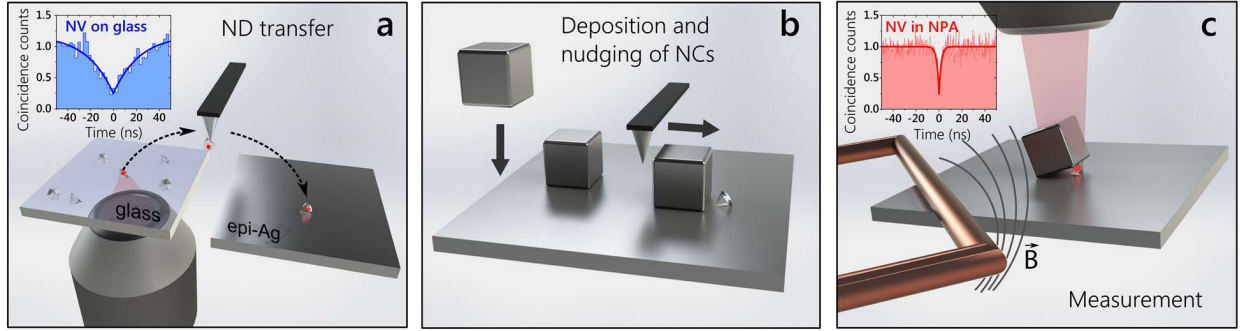

Figure 1. Schematic view of the single-photon NPA assembly and characterization process. (a) Nanodiamonds containing single stable NV centers are identified on a coverslip substrate and transferred to an epitaxial Ag substrate. The inset shows the fluorescence autocorrelation curve indicating the single-photon nature of the emission. (b) Nanocubes are deposited on the epi-Ag substrate and nudged on top of NDs using an AFM tip. (c) Optical characterization using an air objective, with a microwave signal delivered through a copper wire. The inset shows the fluorescence autocorrelation after NPA assembly measured at an excitation rate much below the saturation level.

The emitter preselection on a coverslip substrate aimed at identifying nanodiamonds of suitable size, each containing a single photostable NV center (Fig. 2a). We located the same nanodiamond using an AFM and measure its height (Fig. 2b). Nanodiamonds with the height below 20 nm were sufficiently small for the single-photon NPA assembly and selected for inter-substrate transfer. The candidate nanodiamonds were then picked up by a metal-coated AFM tip, transferred onto the epitaxial Ag substrate (Fig 2c, 2d and 2e), and coated with a 5nm thick $Al_2O_3$ layer. This layer served as the NPA spacer layer, preventing a short-circuit between the cubes and film. After the NCs are randomly deposited on the $Ag/Al_2O_3$ substrate, a cube with approximately 110 nm edge was located and carefully nudged by the AFM tip towards the preselected ND. During this procedure the height of the cube $h_{NC}^0$ only changed by a few nm (Fig. 2f). As the cube was pushed over the ND, its height abruptly changed by the amount close to the ND height $h_{ND}$ (Fig. 2g), evidencing the successful ND-NPA assembly.

After the assembly, we verified the antibunching behavior and noticed that the process allows to preserve the single-photon purity (as illustrated on the insets of Fig. 1a and 1c). At the same time, the antibunching dip narrows down noticeably in the coupled configuration, indicating a significant lifetime shortening, as demonstrated in earlier works utilizing a random assembly process[24,32]. We note that the inter-substrate ND transfer helped to reduce the amount of residue in the assembled NPAs (residues left behind the ND on the glass substrate can be seen on Fig. 2b and 2c). These impurities may have otherwise contributed to the deterioration of the single-photon purity in randomly assembled NPAs.

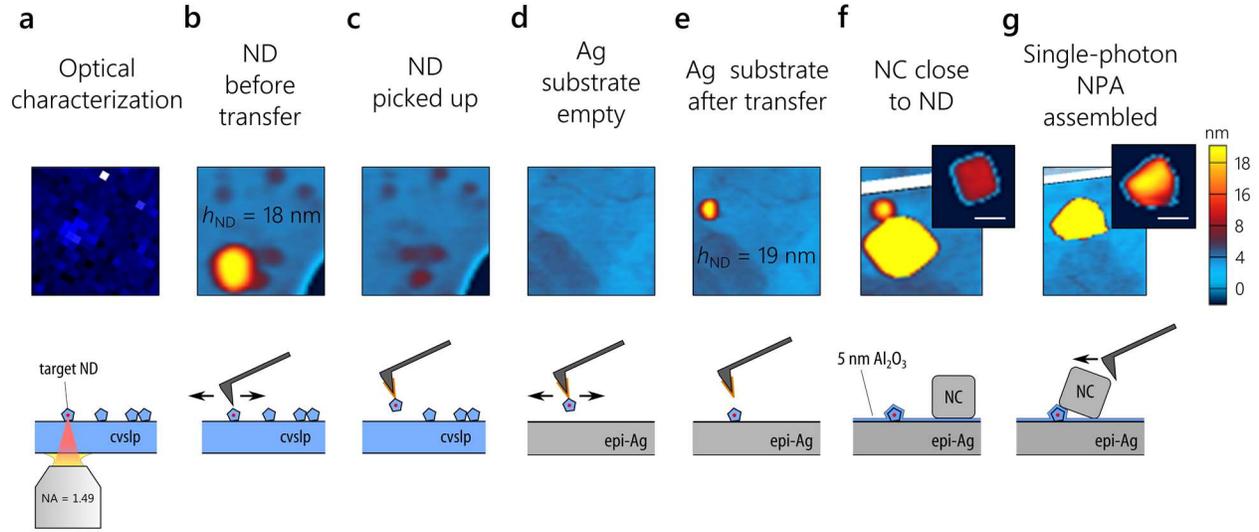

Figure 2. Nanodiamond transfer and single-photon NPA assembly. (a) Optical characterization of a candidate NV on the glass substrate. AFM scans of: (b) the candidate nanodiamond on glass, (c) the same area as in (b) after nanodiamond was picked up, (d) the epi-Ag substrate before nanodiamond placement, (e) the epi-Ag substrate after nanodiamond placement, (f) the nanodiamond on the epi-Ag substrate, covered with a 5 nm alumna layer and a Ag nanocube deposited nearby, (g) the same nanocube pushed on top of the nanodiamond. Insets in (f) and (g) show the same nanocube with a different height scale, illustrating the change in the nanocube height from $h_{NC}^0 = 110$ nm to $h_{NPA} = 126$ nm. The diagrams in the bottom row schematically illustrate the assembly process.

### Optimal fluorescence lifetime shortening:

We then applied the developed approach to assemble an ND-NPA system providing a much larger FLS. NPAs offer a strong FLS in a broad wavelength range, which is highly sensitive to the relative position of the ND and nanocubes (NCs) (Fig. 3a). As the field of the NPA dipolar mode being the strongest at the cube corners (not shown)[33], the strongest FLS is also achieved when the NV is situated at the NC corner.

We simulated the FLS of a nanodiamond-based NV center in an NPA depending on the ND location under the nanocube (Fig. 3a). Out of the 11 simulated configurations, the largest FLS of over 400 times was exhibited by the configuration #5, in which the nanodiamond was positioned under a corner of the nanocube. In this simulation, we assumed a vertical optical dipole (along the z axis), located in the center of a spherical nanodiamond with diameter 15 nm. The nanocube tilt was adjusted for each of the 11 nanodiamond positions to reflect the experimentally realized configurations. The FLS was calculated with respect to a reference configuration with the same nanodiamond positioned on a glass substrate with index $n_{glass} = 1.515$.

In order to realize the optimal FLS scenario, we have assembled a single-photon NPA by placing the nanocube as close as experimentally possible to the configuration #5 from Fig. 3a. In Fig. 3b the AFM scans show the relative positions of the ND with height $h_{ND} = 15$ nm and the NC before and after the coupling. In this image, an appropriate color scale was chosen to illustrate the height change of the NC in the assembly process. Only the contour of the ND is depicted because the particle itself is not rendered on this color scale. Superposing the images of the ND particle before the assembly and that of the assembled NPA, we conclude that the ND particle is located at the NC corner.

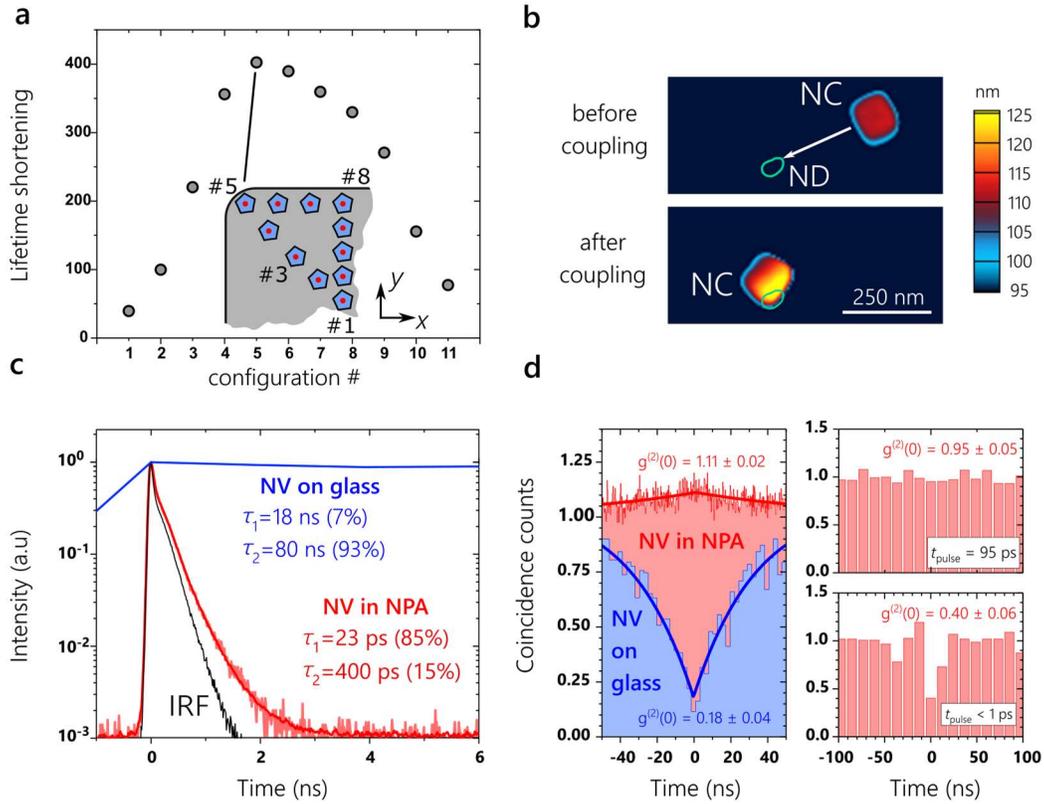

Figure 3. An optimally assembled single-photon NPA. (a) Dependence of the NV FLS on the in-plane position of the ND under the NC. The FLS is calculated relative to an NV in a nanodiamond on glass substrate with index $n_{\text{glass}} = 1.515$ (b) Assembly of the optimal single-photon NPA configuration #5, with the ND situated under a corner of the NC. (c) Comparison between the fluorescence decays of the NV in the uncoupled (blue) and coupled (red) configurations. (d) Autocorrelation measurements of the NV in the uncoupled (blue) and coupled (red) configurations. Under continuous wave (left panel) and long pulse excitation with $t_{\text{pulse}} = 95$ ps (upper right panel) the antibunching cannot be measured. Under short pulse excitation with $t_{\text{pulse}} < 1$ ps (lower right panel), the antibunching appears clearly.

We then characterized the FLS by comparing the fluorescence decays of the NV on the glass substrate and the same NV in the coupled configuration. In both cases, the fluorescence decay exhibits two exponential components, $\tau_1$ and $\tau_2$, with $\tau_1 < \tau_2$. On glass, we obtain $\tau_{1,\text{glass}} = 18 \pm 2$ ns and $\tau_{2,\text{glass}} = 80 \pm 2$ ns, with intensity weights of 7% and 93% respectively. These constants were obtained by a direct two-exponential fit of the measured data. In the NPA configuration, the decay constants are $\tau_{1,\text{NPA}} = 23 \pm 5$ ps and $\tau_{1,\text{NPA}} = 400 \pm 10$ ps, obtained by fitting a two-exponential decay function convoluted with the instrument response function (IRF), plotted in black in Fig. 3c. The intensity weights of these components are 85% and 15% respectively, i.e. the fast decay component dominates the fluorescence. In order to confirm that the fast decay component is indeed due to the NV center, we measure the autocorrelation of the NPA emission using different pump lasers. While the uncoupled NV exhibited a clear antibunching behavior quantified by $g^{(2)}_{\text{NPA}}(0) = 0.18 \pm 0.04$, the NPA emission's antibunching cannot be resolved by the CW autocorrelation measurement at all. Instead, a bunching appears around zero delay, indicating that the emission is not classical. In the pulsed excitation regime, using a laser pulse duration of 95 ps, the antibunching still cannot be resolved, indicating that the fluorescence lifetime is shorter than 50 ps. Finally, using a femtosecond laser, with a pulse duration below 1 ps, the antibunching clearly appears, with $g^{(2)}_{\text{NPA}}(0) = 0.40 \pm 0.06$.

The ratio of the dominant decay components before and after coupling yields an FLS of $\tau_{2,\text{glass}}/\tau_{1,\text{NPA}} = 3500 \pm 800$ times, which is over 8 times larger than FLS predicted in the simulations. This discrepancy may be caused by cavity photo-modification during the pulsed excitation, resulting in a partial sagging of the nanocube and the reduction of the plasmonic cavity volume.

**Spin characterization of the coupled NV centers:**

Yet another strength of the developed deterministic assembly is the ability to choose NV centers with desired properties. NV centers are well-known for their highly coherent spin degree of freedom combined with a spin-dependent fluorescence intensity, allowing an optical spin state readout. The relative difference in fluorescence between the $|m_s = 0\rangle$ and the $|m_s = \pm 1\rangle$ states is termed the spin contrast. In bulk NV centers the spin contrast is a robust phenomenon that withstands saturating optical excitation rates for many days. Conversely, in NVs close to the diamond surface and in particular, in nanodiamond-based NVs, the spin contrast is often reduced compared to its maximal bulk value. In addition, in the majority of our single NV centers, we notice a rapid bleaching of the spin contrast under optical intensities on the order of 1 kW/cm$^2$. However, a careful choice of the nanodiamond candidate particle allowed us to assemble the next nanoantenna with a single photostable NV center possessing also a photostable spin contrast.

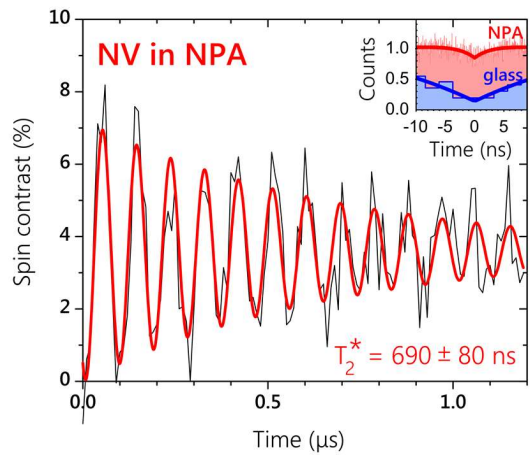

Figure 4. Rabi oscillations detected from a single NV center coupled to a deterministically assembled NPA. The inset shows fluorescence autocorrelation measurements of the NV center in the uncoupled (blue) and coupled (red) configurations.

To characterize the spin properties, we interface the single-photon NPA with a copper wire, connected to an amplified microwave generator. From the Rabi oscillations detected at the frequency of 2.949 GHz, we deduce the inhomogeneous coherence time of $T_{2,\text{NPA}}^* = 690 \pm 80$ ns (Fig. 4). This coherence time is well within the range measured for a set of NVs in nanodiamonds from the same batch ($T_{2,\text{glass}}^* = 700 \pm 200$ ns). We notice that the amplitude of the Rabi oscillations corresponds to a contrast of 7.2% which is well below the maximal intrinsic value of about 30%. By comparing the autocorrelation measurements of the NV on glass and in the NPA (Fig. 4 inset), we find that a significant fluorescence background appeared in the coupled configuration. The background fluorescence percentage can be estimated as $r_{\text{bg, NPA}} = 1 - \sqrt{1 - g_{\text{NPA}}^{(2)}(0)} = 76\%$. This fully explains the observed reduced spin contrast.

## Discussion:

The developed deterministic assembly approach offers significant advantages over random assembly and self-assembly methods. Firstly, it allows to preselect a quantum emitter with a desired set of photophysical, geometric and other properties. Secondly, this method provides a comparison of the photophysical properties of the emitter in the uncoupled and coupled configurations, thus offering a direct assessment of the nanoantenna performance. Provided that nanocubes can also be transported by AFM tips, the NPA assembly could be fully compatible with on-chip integration. By measuring the nanodiamond height and monitoring the evolution of the cube height, we can access important geometrical parameters of the resulting nanopatch antenna. This method can be further extended to control the dipole orientation of the emitter coupled to the NPA.

Previously, it was shown that in the presence of metal the spin relaxation time $T_1$ of the NV spins is reduced due to the magnetic noise from the free electrons in the metal[34]. We show here that the inhomogenous coherence time $T_2^*$ of NV spins in nanodiamonds is not strongly affected by the contact with the metal film. The decoherence rate due to the electrons in metal must therefore not exceed a few MHz, even when the NV is only a few nm from the surface of the epitaxial silver. On a different note, NV ionization is an important process that is believed to limit the magnitude of the spin contrast in both nanodiamond and bulk NVs[35]. The enhancement of the radiative rate in NPAs should prevent the ionization process from erasing the spin contrast at high excitation rates. A sufficient brightness enhancement of NVs in NPAs could lead to the observation of the single-shot electron spin readout at room temperature, provided a sufficient level of single-photon intensity. This in turn opens up new frontiers for applying enhanced NV centers for magnetometry and quantum information processing.

## Methods:

The full-wave 3D electromagnetic numerical simulations were performed using finite-element frequency domain method using the Wave Optics Module of COMSOL Multiphysics. The simulation domain was a 2x3 μm box surrounded by a 400 nm thick perfectly matching layer. The optical emitter was modeled as an AC current density inside a 200 pm diameter sphere enclosed by a 15 nm diameter diamond shell with index $n_{\text{diam}} = 2.42$. The emitter dipole was oriented out-of-plane (along the z axis) and its wavelength was fixed at 680 nm. It was placed on a 60 nm thick epitaxial Ag layer and conformally covered with a 5 nm thick alumina film ($n_{Al2O3} = 1.68$). An Ag nanocube with a 100 nm edge length was placed on top of the nanodiamond. The nanocube corners had an 11 nm curvature radius and the nanocube was conformally covered in a 3 nm thick PVP layer with $n_{PVP} = 1.4$. The bottom facet of the nanocube was tangent to the nanodiamond particle and the edge of the bottom facet furthest from the nanodiamond was resting on the alumina spacer.

The glass sample was prepared by cleaning the coverslip sample with solvents, treating it with ultraviolet radiation for an hour and drying a 5 µL droplet of a sonicated nanodiamond solution (20 nm average size, Adamas Nano) on the coverslip surface. Golden nanowires with 100 nm diameter and 10 µm length (Nanopartz) were dropcast subsequently and used as markers labeling the positions of the nanodiamonds of interest. Surface topography characterization and nanoparticle manipulation were performed using an AFM (Cypher, Asylum Research). Nanodiamond size identification and intersubstrate transfer were performed using Pt coated Si AFM tips (MikroMasch). The silver substrates were prepared by depositing 60 nm epitaxial silver films directly on Si substrate at the BMSTU Nanofabrication Facility (Functional Micro/Nanosystems, FMNS REC, ID 74300). An undiluted water/ethanol solution of 100 nm crystalline nanocubes (Nanocomposix) was dropcasted on the sample. The sample was then rinsed with de-ionized water to remove the PVP residue. Nanocube nudging was performed using bare Si AFM tips (MikroMasch).

All the optical characterization was performed using a custom-made scanning confocal microscope with a 50 µm pinhole based on a commercial inverted microscope body (Nikon Ti−U). Objective scanning was performed using a P-561 piezo stage driven by a E-712 controller (Physik Instrumente). The optical pumping in the CW experiments was administered by a 200 mW continuous wave 532 nm laser (Shanghai Laser Century). Lifetime characterization and picosecond pulsed autocorrelation measurements was performed using a 514 nm fiber-coupled diode laser with a nominal 100 ps pulse width and adjustable repetition rate in the 2−80 MHz range (BDL-514-SMNi, Becker & Hickl). Femtosecond pulsed autocorrelation measurements were performed using a compressed tunable mode-locked laser with a 80 MHz repetition rate (Mai Tai Deep See, Spectra Physics). The laser was set to operate at a wavelength of 1040 nm and its output was frequency doubled to obtain a 520 nm line. The excitation beams in all the optical experiments were reflected off a 550 nm long-pass dichroic mirror (DMLP550L, Thorlabs), and a 550 nm long-pass filter (FEL0550, Thorlabs) was used to filter out the remaining pump power. Two avalanche detectors with a 30 ps time resolution and 35% quantum efficiency at 650 nm (PDM, Micro-Photon Devices) were used for single-photon detection during scanning, lifetime, and autocorrelation measurements. Time-correlated photon counting was performed by an acquisition card with a 4 ps internal jitter (SPC-150, Becker & Hickl).

A single-photon avalanche detector (SPAD) with 69% quantum efficiency at 650 nm (SPCM-AQRH, Excelitas) was used for spin-related measurements. The CW laser was modulated with an acousto-optical modulator (AOM, 1350AF-DIFO-1.0, Gooch & Housego) to produce 5 µs initialization and readout pulses. A pulse counter (USB-CTR04, Measurement Computing) was used for counting the electrical pulses from the SPAD. For the optically detected magnetic resonance (ODMR) and Rabi oscillations measurements, the microwave signal was generated by an Agilent E8254A generator, modulated by a Mini-Circuits ZASW-2-50DR+ switch and amplified by a 16 W Mini-Circuits ZHL-16W-43 RF amplifier. The AOM, the counter, and the microwave switch were gated using a delay generator (QC9520, Quantum Composers), producing appropriate pulse sequences.

**Author Contributions**

**Acknowledgements**

This work was partially supported by the U.S. Department of Energy, Office of Basic Energy Sciences, Division of Materials Sciences and Engineering under Award DE-SC0017717 (supplies and personnel) and the Office of Naval Research (ONR) DURIP Grant No. N00014-16-1-2767 (equipment grant used to purchase the scanning confocal microscope, lasers, detectors and single-photon counting capability used in this work). A.V.K. acknowledges the DARPA/DSO EXTREME, Award HR00111720032 (numerical modeling and simulations).

**Supporting Information**

Details about the simulation of the emitters' decay rates, statistics of reference emitter characteristics, data for additional single-photon nanoantennas, description of the nanoantenna photomodification experiment and simulation of the photomodified configuration.